\documentclass[12pt]{article}
\usepackage{amsmath,cite}
\usepackage{amssymb}
\usepackage{slashed}
\usepackage[hidelinks]{hyperref}
\usepackage{cite}
\newcommand{\p}[1]{(\ref{#1})}
\newcommand{\be}{\begin{equation}}
\newcommand{\ee}{\end{equation}}
\newcommand{\bea}{\begin{eqnarray}}
\newcommand{\eea}{\end{eqnarray}}
\newcommand{\pl}{\partial}

\def\theequation{\arabic{section}.\arabic{equation}}
\begin{document}
\topmargin -1cm \oddsidemargin=0.25cm\evensidemargin=0.25cm
\setcounter{page}0
\renewcommand{\thefootnote}{\fnsymbol{footnote}} 
\begin{titlepage}
\vskip .7in
\begin{center}
{\Large \bf  Massive Higher Spins and Black Hole Interactions
 } \vskip .7in 
 {\Large 
Julian Lang\footnote{e-mail: {\tt  julian.lang@oist.jp }} and 
 Mirian Tsulaia\footnote{e-mail: {\tt  mirian.tsulaia@oist.jp}  }}
 \vskip .4in{  \it Okinawa Institute of Science and Technology, \\ 1919-1 Tancha, Onna-son, Okinawa 904-0495, Japan}\\
\vskip .8in
\begin{abstract}

We give a  brief introduction into the gauge invariant formulation of irreducible massive bosonic 
higher spin fields. We discuss both free Lagrangians
and the ones which include  cubic interactions. We demonstrate an application of these Lagrangians to a description of the interactions between Kerr black holes in a Post Minkowskian approximation.

\end{abstract}

\end{center}

\vfill

\end{titlepage}

\renewcommand{\thefootnote}{\arabic{footnote}}
\setcounter{footnote}0

\newpage

\section{Introduction}\setcounter{equation}0

One of the many possible physical systems, where Gravitational Waves
(GW)
occur, is the interacting  black hole binary.
A direct use of the Einstein equations for the description
of such a system is complicated, and therefore one has to resort to other methods.
For example, one can  use an Effective Field Theory approach 
\cite{Cangemi:2022bew}
and  model rotating (Kerr) black holes as localized massive objects (particles) with a  spin. 
This approximation  is possible,
due to   the "No - hair" theorem and 
 when the size of black hole 
is much smaller than the wavelength of the Gravitational Waves. 
Below we shall   consider
 a Post Minkowskian Approximation i.e., weak fields
$\frac{Gm}{rc^2}<<\frac{v^2}{c^2}$
but high velocities $\frac{v^2}{c^2} \sim 1$, and the 
perturbation is performed around special relativity, i.e., we have an expansion in $G$
(Newton constant).

The energy momentum tensor for a Kerr black hole
in the Post Minkowskian approximation was found in the world-line formalism  \cite{Vines:2017hyw}
to be\footnote{For any two $A_\mu$ and $B_\mu$
we denote
$A \cdot B \equiv A^\mu B_\mu$.}
\be
T^{\mu \nu}(-k) = 2 \pi \delta (p \cdot k) p^{(\mu}
\exp\left (
\frac{S * ik}{m} \right)^{\nu)}_\rho p^\rho, \quad
 (S * ik)^\mu{}_\nu = \varepsilon^{\mu}{}_{\nu  \rho \sigma} S^\rho k^\sigma\ ,
\ee
 where $p$ is the black hole momentum, $k$ the GW momentum, and $S^\mu$ is a spin operator.
Using this expression for the energy momentum tensor one can  write a cubic  interaction 
with a  graviton, described by a polarization tensor
$\varepsilon^{\mu \nu}(k) =\varepsilon^{\mu}(k) \varepsilon^{\nu}(k)$ as
\be \label{K-GW}
V_{3.gr} =\varepsilon^{\mu}(k) \varepsilon^{\nu}(k)  T^{\mu \nu}(-k).
\ee
This interaction can also be obtained from the field theory
approach in the following way.

Consider a three point on-shell scattering amplitude
 between two fields with the same mass $m$ and an arbitrary spin $s$ (labels $`1`$ and $`2`$), with a massless field with a helicity $h$
 (label $`3`$). 
 Using the four dimensional spinor helicity formalism
 \cite{Dittmaier:1998nn}--\cite{Chiodaroli:2021eug}  one can show
 that
 this three point amplitude has the general form \cite{Chung:2018kqs}
 \be
M_3^{s,s,h} = 
(mx)^h \left (
g_0 \frac{\langle 2 1 \rangle^{2s}}{m^{2s}} +
g_1 x \frac{\langle 2 1 \rangle^{2s-1} \langle 2 3 \rangle \langle 3 1 \rangle
}{m^{2s+1}}
+...\right),
\ee
where $g_t$ are coupling constants, and the factor $x$ is defined as
\be \label{001}
\frac{\sqrt 2}{m} \epsilon^{(3),+} \cdot p^{(1)} =
x, \quad 
\frac{\sqrt 2}{m} \epsilon^{(3),-} \cdot p^{(1)} =
\frac{1}{x}.
\ee
If all coupling constants  except $g_0$ are zero, then the three point amplitude describes the  minimal coupling
\cite{Arkani-Hamed:2017jhn}. For the case of 
interaction with Gravity ($h=2$) or with the vector field
($h=1$) one therefore, has
\be \label{min-c}
M_3^{s,s,2} =  (mx)^2 
 \frac{\langle 2 1 \rangle^{2s}}{m^{2s}}, \quad
M_3^{s,s,1} =  (mx) 
 \frac{\langle 2 1 \rangle^{2s}}{m^{2s}}.
\ee
Finally, by carefully taking  the classical limit
\cite{Guevara:2018wpp},
\cite{Chung:2018kqs},
\cite{Chiodaroli:2021eug}
of the three point amplitude 
\p{min-c} one obtains the interaction of the 
Kerr black hole with GW, given in 
\p{K-GW}.

Our goal is to show how to construct an off-shell
extension of the three point amplitude 
\p{min-c} by using a BRST invariant
formulation of massless and massive higher spin fields
\cite{Skvortsov:2023jbn}. This approach closely follows
the one of String Field Theory
(see also \cite{Alessio:2025nzd} and 
\cite{Ponomarev:2022vjb} -- \cite{Tran:2020uqx} for recent reviews on higher spin theories)
and will be described in the following Sections.

\section{Free massive higher spin fields}
Below we shall follow a gauge invariant BRST approach
\cite{Pashnev:1997rm}--\cite{Buchbinder:2005ua},
(see \cite{Pashnev:1989gm}--
\cite{Fegebank:2024yft} for different Lagrangian formulations for  massive higher spin fields).
 Higher spin fields are usually described by rank $s$ symmetric tensors. 
Tensors can be traceless or not, therefore representing irreducible or reducible representations of the Poincar\`e
 (or $AdS_D$, $dS_D$) group.
 In addition, these
fields must satisfy the mass-shell and transversality conditions.
Therefore, reducible massive bosonic fields
on a flat background must satisfy equations\footnote{Reducible massless higher spin multiplets were considered in the frame-like formulation in \cite{Sorokin:2008tf}.}
\begin{eqnarray} \label{ms}
&&(\Box - m^2) \phi_{\mu_1 \mu_2,...,\mu_s}(x)=0, \quad \text{mass-shell} \\ \label{transv}
&&\partial^\mu \phi_{\mu \mu_2,...,\mu_s}(x)=0. \quad \,\,\,\, \,\,\,\,\,\, \,\,\,\,\,\,\,\,\,\,\,\text{transverse} 
\end{eqnarray}
Below we shall consider reducible representations,
since interaction vertices for them have a much simpler form, in comparison to irreducible fields.

\subsection{An example: massive scalar relativistic particle}

Our task is to construct a quadratic action, which gives equations \p{ms} -- \p{transv}
as equations of motion.
 In order to illustrate the main idea of this construction,
let us consider an example of a scalar relativistic particle. Its worldline  $x^\mu(\tau)$ is parametrized by the proper time $\tau$ and the corresponding
 action is
\be \label{act-sc-1}
S = - m \,\int d \tau \sqrt{- \dot{x}^\mu(\tau) \, \dot{x}_\mu(\tau) },
\quad  \dot{x}^\mu(\tau) = \frac{d x^\mu}{d \tau}.
\ee
The canonical momentum
\be
p_\mu  =
\frac{m \dot{x}_\mu}{\sqrt{- \dot{x}^\nu \, \dot{x}_\nu}}
\ee
obeys the constraint
\be
l_0 \equiv (p \cdot p + m^2)=0 
\ee
as a result of the reparametrization invariance
$\tau \rightarrow  \tau(\tau^\prime)$
of the action. Introducing an anticommuting nilpotent ghost variable $c_0$ one builds  a BRST charge
\be
Q = c_0 l_0 = c_0 (p^\mu \, p_\mu + m^2); \quad c_0^2 =0,
\ee
which according to the method of BRST quantization is imposed on a physical state
\be
c_0(p \cdot p + m^2)| Phys \rangle = 0.
\ee
This equation clearly implies that the physical state 
satisfies the massive Klein -- Gordon equation.
In order to build a field theoretic Lagrangian that gives 
it as an equation of motion
\cite{Siegel:1984ogw}
we write
\be
|Phys \rangle = \phi(x) | 0 \rangle 
\ee
and consider a quadratic Lagrangian
\be \label{act-sc-2}
{\cal L} = - \int d c_0 \, \langle Phys | Q | Phys \rangle
= - \phi(x) \, (p^2 + m^2) \phi(x),
\ee
where we have performed the integration over the Grassmann variable $c_0$.  

\subsection{BRST construction for free massive higher spin fields}

Now we would like to apply the same method for reducible
massive higher spin fields. Let us note, that unlike
the case of the relativistic particle, we
do not consider any mechanical model that gives
the equations \p{ms} -- \p{transv} as a result of the Hamiltonian analysis. Rather, we impose them axiomatically.

To simplify the description of the tensor fields, we
 introduce an auxiliary Fock space spanned by oscillators
 \be
[\alpha_\mu, \alpha_\nu^+] = \eta_{\mu \nu}, \quad
[\xi, \xi^+]=1, \quad \alpha_\mu |0 \rangle=
\xi |0 \rangle =0, 
\ee
and we take mostly plus signature for the metric.
For mass-shell, modified  divergence and modified gradient operators we have
\be \label{oper}
l_0 = p \cdot p + m^2, \quad l = \alpha \cdot p + m \xi,
\quad l^+ =    \alpha^+ \cdot p+ m \xi^+,
  \quad  \quad p_\mu = -i \partial_\mu.
\ee
We use the term ``modified" because of the presence of the term, proportional to the mass $m$. At the end, after gauge fixing, these operators will effectively  reduce
to standard divergence and gradient operators, as we shall
see below.
The  
operators \p{oper} form a closed algebra,
the only nonzero commutator being
\be
[l, l^+]=l_0.
\ee
After introducing Grassmann-odd  ghost variables, 
with only nonzero anticommutation relations being 
\be
 \{ c_0, b_0 \} = \{ c, b^+ \} = \{c^+, b \} =1,
\quad c|0 \rangle = b|0 \rangle= b_0|0 \rangle=  0,
 \ee
one builds a nilpotent BRST 
operator in a standard way
\be \label{BRSTcharge}
Q= c_0 l_0 + c^+l + c l^+ - c^+ c \, b_0, \quad Q^2=0  .
\ee
A vector in this enlarged Fock space has the form
\be
 |\Phi \rangle = | \varphi^{(s)} \rangle + c_0 b^+ | C^{(s-1)} \rangle + c^+ b^+ |D^{(s-2)} \rangle .
 \ee
 Where the vectors $| \varphi^{(s)} \rangle$, $|C^{(s-1)} \rangle$, and $|D^{(s-2)} \rangle $ are expanded in terms of only
$\alpha^+_\mu$ and $\xi^+$ oscillators with total numbers $s$,
$s-1$, and $s-2$. For example
\be \label{vec-exp}
| \varphi^{(s)}  \rangle = \sum_{k=0}^{k=s}\frac{1}{(s-k)! k!} \varphi_{\mu_1 \mu_2,...\mu_{s-k}}(x) \alpha^{ \mu_1,+} \alpha^{ \mu_2,+}
...\alpha^{\mu_s,+} (\xi^+)^{k}| 0 \rangle.
\ee
 Now, in analogy with the relativistic particle model
considered above we write a quadratic Lagrangian
\be \label{L-BRST}
{\cal L} = - \int d c_0  \langle \Phi|Q | \Phi \rangle. 
\ee
Due to nilpotency of the BRST charge, the Lagrangian
is invariant under the gauge transformations
\be \label{GT-BRST}
\delta | \Phi \rangle = Q | \Lambda \rangle, \quad
| \Lambda \rangle = b^+ | \lambda^{(s-1)} \rangle,
\ee
where the vector $| \lambda^{(s-1)} \rangle$ is expanded
only in terms of oscillators $\alpha^{+}_\mu$ and
$\xi^{+}$ similarly to \p{vec-exp}.
Performing the Grassmann integration over $c_0$ 
and  normal ordering with respect to the other ghost
variables, one obtains
the Lagrangian \cite{Pashnev:1997rm}
\begin{eqnarray} \label{Lagr-comp}
{\cal L} &=& - \langle \varphi^{(s)}| (p \cdot p + m^2) | \varphi^{(s)} \rangle +
\langle D^{(s-2)}|( p \cdot p + m^2)| D^{(s-2)} \rangle
\\ \nonumber
&+& \langle \varphi^{(s)}|(\alpha \cdot p + m \xi)| C^{(s-1)} \rangle -
\langle C^{(s-1)}|( \alpha \cdot p + m \xi)| D^{(s-2)} \rangle
  \\ \nonumber
&+& \langle C^{(s-1)} |(\alpha^+ \cdot p + m \xi^+ )|  \varphi^{(s)} \rangle -
\langle D^{(s-2)} |(\alpha^+ \cdot p + m \xi^+ )| 
 C^{(s-1)} \rangle
\\ \nonumber  
&-&  \langle C^{(s-1)}|  | C^{(s-1)} \rangle ,
\end{eqnarray}
which is invariant under the gauge transformations
\begin{eqnarray} \label{GT-b}
&&\delta |\varphi^{(s)} \rangle= (\alpha^+ \cdot p + m \xi^+)
|\lambda^{{(s-1)}} \rangle, \\ \nonumber
&&\delta |C^{(s-1)} \rangle= (p \cdot p + m^2 )
|\lambda^{{(s-1)}} \rangle, \\ \nonumber
&&\delta |D^{(s-2)} \rangle= (\alpha \cdot p + m \xi)
|\lambda^{{(s-1)}} \rangle,
\end{eqnarray}
where we used the equations
\p{BRSTcharge}, \p{vec-exp}
and \p{GT-BRST}. 
One can prove, that after gauge fixing, one is left  only with the physical field $| \varphi^{(s)} \rangle$, 
with no $\xi^+$ dependence.
The rest of the fields are either pure gauge, or zero due to the equations 
of motion. The physical field then
 satisfies
\be
(p \cdot p + m^2) | \varphi^{(s)} \rangle= \alpha \cdot p | \varphi^{(s)} \rangle=0.
\ee
 Let us note, that in order to obtain 
an analogous description for massless reducible higher spin fields, one can simply put $m=0$ and discard the $\xi^\pm$ dependence everywhere.

\subsection{An example: free massive spin one field}
As an illustration of this technique, we give an example
of a gauge invariant formulation of a massive spin one 
field.
Since in this case $s=1$, the fields 
\p{vec-exp}
and the parameters of gauge transformations
\p{GT-BRST} take the form
\be
| \Phi \rangle = | \varphi^{(1)} \rangle  + c_0 b^+ | C^{(0)} \rangle, \quad
| \Lambda \rangle = ib^+ \lambda (x ) | 0 \rangle,
\ee
 where
\be
| \varphi^{(1)} \rangle=  \left ( \varphi_{\mu} (x) \alpha^{\mu +} + i
\varphi (x) \xi^{ +} \right )| 0 \rangle, \quad
| C^{(0)} \rangle =
-i  C(x)  | 0 \rangle.
\ee
After performing the normal ordering of
the $\alpha$ and $\xi$ oscillators, 
the Lagrangian \p{Lagr-comp}
takes the form
\be \label{lagr-1}
{\cal L}=   \phi^{\mu} (\Box - m^2)  \phi_{\mu}
+
\phi (\Box - m^2)  \phi -  C^2
+ 2 C \partial^\mu \phi_{\mu} - 2 m C \phi.
\ee
The gauge transformations 
\p{GT-b} 
are
\be
\delta \phi_{\mu} (x) = \partial_\mu \lambda(x), \quad
\delta \phi (x) = m \lambda(x), \quad
\delta C (x) = (\Box-m^2) \lambda(x).
\ee
From the Lagrangian \p{lagr-1}
one can see, that 
after expressing the field $C(x)$ in terms
of $\phi_{\mu} (x) $ and $\phi(x) $
by using its own equation of motion, one obtains
a gauge invariant (Stueckelberg)
formulation for a massive vector
field. Furthermore, one can gauge away the field $\phi(x)$ and obtain a massive and transverse
vector field $\phi_{\mu} (x)$. Alternatively,
one can gauge away the time-like component
of the field $\phi_{\mu} (x)$. Then the field 
$\phi (x)$ will correspond to a physical degree of freedom of the massive vector field.

\section{Diagonalization Procedure}
In the above, we considered free Lagrangians for reducible representations of the
Poincar\'e group.
 A natural question that arises 
is, whether one can decompose these Lagrangians into a sum of Lagrangians for each individual
irreducible representation.
  We shall refer to this procedure as the diagonalization procedure.
Up to now, the diagonalization procedure has been considered only for massless representations \cite{Fotopoulos:2009iw}--
\cite{Glennon:2025yjp}. 
Presumably iagonalization for massive representations
 can be performed along the same lines and
is an interesting open problem.

Let us explain the diagonalization procedure on the
examples of $s=2$
and $s=3$.
After excluding the field $C_\mu(x)$
via its own equation of motion, the Lagrangian \p{Lagr-comp}
for the case $s=2$ and $m=0$ reads
\be \label{L20}
{\cal L}=  \frac{1}{2} \phi^{\mu \nu} \Box \phi_{\mu \nu} -
2D \Box D 
+2  \phi_{\mu \nu} \partial^\mu \partial^\nu D 
- \phi^{\tau \nu} \partial_\tau \partial^\mu 
\phi_{\mu \nu}  
\ee
and is invariant under gauge transformations
\be
\delta \phi_{\mu \nu} = \pl_\mu \lambda_\nu
+  \pl_\nu \lambda_\mu, \quad \delta D=
\pl^\mu \lambda_\mu.
\ee
Introducing a gauge invariant scalar $\Phi$
and a field $\Phi_{\mu \nu}$ according
to
\be
\Phi =  \phi^{\mu}{}_{\mu} -2 D,
\quad  \phi_{\mu \nu}=  \Phi_{\mu \nu} + \frac{1}{d-2}\eta_{\mu \nu} \Phi,
\ee
and substituting into the Lagrangian
\p{L20}, one can see, that it splits into a sum of two pieces:
one describing an irreducible field with spin $2$
(a linearized graviton) and the other describing a scalar field
\bea \nonumber \label{2+0}
{\cal L} &=&  \frac{1}{2}\Phi^{\mu \nu} \, \Box \, \Phi_{\mu \nu} - \Phi^{\mu \tau} \partial_\tau \partial^\rho
\Phi_{\mu \rho}
-\frac{1}{2}\Phi^{\mu}{}_{\mu} \, \Box \, \Phi^{\nu}{}_{\nu}+
\Phi^{\rho}{}_{\rho} \, \partial_\mu \partial_\nu \, \Phi^{\mu \nu} \\ 
&+& \frac{1}{2(d-2)} \Phi \, \Box \, \Phi.
\eea
The transformations that diagonalize the Lagrangian
\p{Lagr-comp} 
for $s=3$ with $m=0$, with the auxiliary field
$C_{\mu \nu}(x)$
being excluded via its own equation of motion,
are
\be \label{db3}
 \phi_{ \mu_1 \mu_2 \mu_3} = \Phi_{ \mu_1 \mu_2 \mu_3} + \frac{1}{d}  \eta_{(\mu_1 \mu_2} \Phi_{\mu_3)},
\quad
\Phi_\mu =  \phi^\nu{}_{\nu \mu} - 2  D_\mu.
\ee
Alternatively, these transformations can be obtained
 from the requirement that
  the fields
$\Phi_{ \mu_1 \mu_2 \mu_3}$ and $\Phi_{ \mu}$ have 
``proper" gauge transformation rules i.e., 
\be
\delta \Phi_{ \mu_1 \mu_2 \mu_3}= \partial_{(\mu_1 } {\tilde \lambda}_{\mu_2 \mu_3)},
\quad
\delta {\Phi}_{ \mu}= \partial_{\mu } \tilde \lambda,
\ee
where the parameter  $ \tilde \lambda_{\mu \nu}$ is traceless.
General cases  of an arbitrary  $s$ for bosonic, and for fermionic  totally
symmetric  massless representations the Poincar\'e group (Young  diagrams of type $Y(s,0)$) ,
as well as for  bosonic massless representations with mixed symmetry
of type $Y(s-1,1)$, were considered in \cite{Fotopoulos:2009iw}--\cite{Glennon:2025yjp}.

\section{Cubic Interactions}

Finally, we are ready to add cubic interactions
between the fields considered in the previous section.
Keeping  in mind applications
to rotating Black Holes, we consider cubic interactions between 
two  fields with equal mass $m$ and arbitrary spin $s$, and one  massless field with spin two (or spin one, if one wants to consider root-Kerr which is useful for double-copy).

Let us first describe the general formalism.
We take three copies of the oscillators considered in the previous section. These copies are labeled by the index $(i)$. It takes values $i=1,2$ for massive fields and $i=3$ for massless fields. Therefore,
\begin{eqnarray} \label{osc}
&&[\alpha_\mu^{(i)}, \alpha_\nu^{{(j)},+}] = \eta_{\mu \nu} \delta^{ij}, \quad i=1,2,3;
\\ \nonumber
&&
[\xi^{(1)}, \xi^{{(1)},+}] =
[\xi^{(2)}, \xi^{{(2)},+}]=
1.
\end{eqnarray}
To describe nontrivial cubic interactions,
we need nonlinear gauge transformations
\bea \label{n-ll}
&&\delta | \Phi^{(i)} \rangle = Q^{(i)}  |\Lambda^{(i)} \rangle -
\\ \nonumber
&&- \int dc_0^{i+1} dc_0^{i+2}
g(\langle \Phi^{(i+1)}| \langle \Lambda^{(i+2)}| +  \langle \Lambda^{(i+2)}|
\langle \Phi^{(i+1)}|
) | V_3 \rangle ),
\eea
where $g$ is a coupling constant and no summation over the repeated indexes is assumed.
The corresponding cubic Lagrangian is
\be \label{n-lgt}
{\cal L}_{cub.} = - \sum_{i=1}^3\int dc_0^{i} \langle \Phi^{(i)} |Q^{(i)}  | \Phi^{(i)}  \rangle - g\left(\left(
\prod_{i=1}^3 \int dc_0^{i}
\langle \Phi^{(i)} |
\right)|V_3 \rangle + h.c.\right).
\ee
The cubic vertex has the structure
$
|V_3 \rangle = V_3 c_0^1 c_0^2  c_0^3 |0 \rangle
$
where $V_3$ is an unknown function, to be determined from the requirement of 
invariance of the Lagrangian 
\p{n-ll}
under the transformations
\p{n-lgt}.
The invariance in the zeroth power
of the coupling constant is  fulfilled
due to the nilpotency property of the BRST charges
$(Q^{(i)})^2=0$. 
The invariance in the fist power of $g$ implies that the vertex is BRST invariant
\be \label{cubic-v}
(Q^{(1)} + Q^{(2)} + Q^{(3)} ) | V_3 \rangle=0.
\ee
Notice that the same condition appears as the requirement for the closure of the algebra
of gauge transformations up to the first order in $g$.

Up to now as our
 approach is general, it can be 
 applied for any physical system
 which is gauge invariant up to the cubic order of interactions. The choice of the system is tantamount to the choice of $Q^{(i)}$ and of $ | \Phi^{(i)}  \rangle$.

Let us now address the problem at hand.  
Complete classification of the BRST invariant cubic vertices on a flat background was obtained in
\cite{Metsaev:2012uy}
(See\cite{Metsaev:2005ar}--\cite{Metsaev:2022yvb} for the corresponding vertices in the light-cone approach
and
\cite{Buchbinder:2021xbk} for irreducible massless fields).
 In particular for the case of two massive fields with the same mass and one massless field,
we have BRST invariant combinations
\begin{align}
		\mathcal{K}^{(1)}&=(p^{(2)}-p^{(3)})\cdot \alpha^{(1), +} - m
        \xi^{{(i)},+} 
        +(b_0^{(2)}-b_0^{(3)})c^{(1), +}\ ,\\
		   \mathcal{K}^{(2)}&=(p^{(3)}-p^{(1)})\cdot \alpha^{(2), +} +m
        \xi^{{(2)},+}
        +(b_0^{(3)}-b_0^{(1)})c^{(2), +}   ,\\
		  \mathcal{K}^{(3)}&=(p^{(1)}-p^{(2)})\cdot \alpha^{(3), +}
        +(b_0^{(1)}-b_0^{(2)})c^{(3), +} ,\\
		\mathcal{Q}&=\mathcal{Q}^{(1,2)}+
        \frac{\xi^{{(1)},+}}{2m}\mathcal{K}^{(2)}-\frac{\xi^{{(2)},+}}{2m}\mathcal{K}^{(1)}\ ,\\
		\mathcal{Z}&=\mathcal{Q}^{(1,2)}\mathcal{K}^{(3)}+\mathcal{Q}^{(2,3)}\mathcal{K}^{(1)}+\mathcal{Q}^{(3,1)}\mathcal{K}^{(2)}\ ,
	\end{align}
	where
	\begin{align} \nonumber
		Q^{(1,2)}&=\alpha^{(1), +}\cdot
\alpha^{(2), +}+\xi^{{(1)},+} \xi^{{(2)},+}+
\frac{1}{2}\left(b^{(1 ),+} c^{(2 ),+} + 
b^{(2 ),+} c^{(1 ),+}
\right)\ ,\\
		Q^{(i,i+1)}&=
        \alpha^{(i), +}\cdot
\alpha^{(i+1), +}+\frac{1}{2}\left(b^{(i ),+} c^{(i+1 ),+} + 
b^{(i+1 ),+} c^{(i ),+}
\right)\,      
        \end{align}
for $i=2,3$.

\section{Quartic Interactions}
One can continue to quartic interactions
along the same lines as was done
for the case of massless fields in \cite{Taronna:2011kt}--\cite{Karapetyan:2021wdc}. In particular,
one introduces four sets of oscillators  \p{osc} and adds to the Lagrangian \p{n-lgt} quartic interaction terms
\be
g^2\sum_{a}\left(\left(
\prod_{i=1}^4 \int dc_0^{i}
\langle \Phi^{(i), (a)} |
  | \right)|W^{(a)} \rangle + h.c.\right)
\ee
where $(a)$ indicates different possibilities
for how massless and massive fields can be distributed between the four Fock spaces. Similarly, the gauge transformations
\p{n-ll}
also get modified by terms proportional to $g^2$. The quartic vertices are then determined by the requirement of gauge invariance and closure of the algebra of gauge transformations (see \cite{toappear} for details).

\section{Cubic action for Black Holes}  

A priori, any function of the BRST invariant
combinations ${\mathcal K}^{(i)}$, ${\mathcal Q}$ and ${\mathcal Z}$,
is a valid cubic vertex, since overall coupling 
constants are not fixed by the BRST invariance condition
\p{cubic-v}.
Our task is to find a function that gives minimal coupling, when
the states entering the cubic vertex are regarded as on-shell states. 
Using the spinor-helicity formalism
one can see, that $\mathcal{K}^{(3)}$
and $\mathcal{Z}$ gives \p{min-c} 
 for $s=0$, $h=1$ and $s=1$, $h=1$.
The general solution can be obtained by
 extending an on-shell result
of \cite{Chiodaroli:2021eug} (see \cite{Skvortsov:2023jbn} for details).
For the case of $s-s-2$ we obtain the vertex
\be
V(s,s,2)= ({\cal K}^{(3)})^2
+
\left ( {\cal Z}{\cal K}^{(3)}+  \frac{{\cal Z}^2 -
{\cal Q}^2 {\cal K}^{(3)} {\cal Z}
}{ (1 - {\cal Q})^2  - \frac{1}{2m^2} 
{\cal K}^{(1)} {\cal K}^{(2)}
}
\right).
\ee
For the case $s-s-1$, one has
\be
V(s,s,1)= 
{\cal K}^{(3)}+\frac{{\cal Z} -
{\cal Q}^2 {\cal K}^{(3)}
}{ (1 - {\cal Q})^2  - \frac{1}{2m^2} 
{\cal K}^{(1)} {\cal K}^{(2)}
}.
\ee
Thus  using a gauge invariant BRST approach to massive and massless higher spin fields
one can produce a Lagrangian description 
of the three point amplitude given in \cite{Arkani-Hamed:2017jhn}.

\section*{Acknowledgements}  
We are grateful to  K. Glennon, and  E. Skvortsov  for collaboration on the topics presented
in this paper and to Y. Neiman for useful discussions.
The work  was supported by the Quantum Gravity Unit of the Okinawa Institute of Science and Technology Graduate University (OIST).

\renewcommand{\thesection}{A}

\renewcommand{\theequation}{A.\arabic{equation}}

\setcounter{equation}0
\appendix
\numberwithin{equation}{section}

\section{Spinor Helicity Formalism.  Definitions}\label{Appendix A}

Here we collect some definitions for the four dimensional
massive spinor - helicity formalism (for more identities see for example \cite{Chung:2018kqs}, \cite{Chiodaroli:2021eug}--\cite{Skvortsov:2023jbn}).

 A four - momentum for a massive particle with $p \cdot p =-m^2$ is parametrized
by Dirac spinors
 $\lambda_\alpha^a$
and $\bar \lambda_{\dot \alpha}^a$, with $a =1,2$
being $SU(2)$ group index
\be
p_{\alpha \dot \alpha} = \lambda_\alpha^a \bar \lambda_{\dot \alpha,a}, \quad p_{\alpha \dot \alpha} =  (\sigma^\mu)_{\alpha \dot \alpha} \, p_\mu,
\ee
$$
(\sigma^\mu)_{\alpha \dot \alpha} ({\bar \sigma}^\nu)^{\alpha \dot \alpha} = -2 \eta^{\mu \nu}, \quad
(\sigma^\mu)_{\alpha \dot \alpha} ({\bar \sigma}_\mu)^{\beta \dot \beta} = -2 \delta^\alpha_\beta  \, \delta^{\dot \alpha}_{\dot \beta}. 
$$
The spinors and the sigma matrices are normalized as 
\be 
 \lambda^{\alpha, a} \lambda_\alpha^b = -m \, \epsilon^{ab}, \quad
{\bar \lambda}_{{\dot \alpha}, a} {\bar \lambda}_b^{\dot \alpha} = - m \,\epsilon_{ab}.
\ee
It is convenient   to introduce  auxiliary commuting variables
$z_a$ and consider  spinors with the $SU(2)$ indices being suppressed
$
{\lambda}_\alpha = \lambda_\alpha^a z_a$ and $
{ {\bar \lambda}_{\dot \alpha}} = {\bar \lambda}_{\dot \alpha}^a z_a.
$
 The massive spinors satisfy the Dirac equation
\be
p_{\alpha \dot \alpha } {\bar \lambda}^{\dot \alpha}
= m \lambda_\alpha, \quad
p_{\alpha \dot \alpha }  \lambda^{ \alpha}
= -
m {\bar \lambda}_{\dot \alpha}.
\ee
For massless particles  $p \cdot p =0$, one has simply
$
p_{\alpha \dot \alpha} = \lambda_\alpha \bar \lambda_{\dot \alpha}
$
where $\lambda_\alpha$ and $\bar \lambda_{\dot \alpha}$
are Weyl spinors without internal indices.
It is also convenient to introduce bra- and ket- notations 
\be
| \lambda \rangle \leftrightarrow \lambda_\alpha, \quad
| \lambda ] \leftrightarrow {\bar \lambda}^{ \dot \alpha}
\quad
\langle \lambda |  \leftrightarrow \lambda^\alpha, \quad
[ \lambda |  \leftrightarrow {\bar \lambda}_{ \dot \alpha}
\ee
\be
\lambda^\alpha = \epsilon^{\alpha \beta} \lambda_\beta,\quad
\lambda_\alpha = \epsilon_{\alpha \beta} \lambda^\beta, \quad
{\bar \lambda}^{\dot \alpha} = \epsilon^{{\dot \alpha} {\dot  \beta}} {\bar \lambda}_{\dot \beta}, \quad
{\bar \lambda}_{\dot \alpha} = \epsilon_{{\dot \alpha} {\dot  \beta}} {\bar \lambda}^{\dot \beta}
\ee
and 
\be
\langle i j \rangle = \lambda^{(i), \alpha} \lambda^{(i)}_ {\alpha}, \quad [ i j ] = {\bar \lambda}^{(i)}_{\dot \alpha} 
{\bar \lambda}^{(i), \dot \alpha}, 
\ee
where $i$ and $j$ label the particles.
For  vector fields we have polarization vectors
\be
\epsilon_{m \neq 0} = \sqrt 2\frac{| \lambda \rangle [\lambda |}{m},
\quad
\epsilon^+_{m = 0} = \sqrt 2\frac{| q \rangle [\lambda |}{ \langle q \lambda \rangle},
\quad
\epsilon^-_{m = 0} = \sqrt 2\frac{| \lambda \rangle [q |}{[\lambda q]},
\ee
where $|q \rangle$ and $|q ]$ are arbitrary reference spinors.
 For an arbitrary spin $s$ field the polarization tensor
is
$
\epsilon^{(s)} = (\epsilon)^s.
$

\end{document}